\documentclass{ws-mpla}
\usepackage[super]{cite}
\usepackage{graphicx}
\usepackage{amsmath}
\usepackage{bm}

\begin{document}

%%%%%%%%%%%%%%%%%%%%%%%%%%%%%%%%%%%%%%%%%%%%%%%%%%%%%%%%%%%%%%%%%%%%

\title{On the one-loop calculations with Reggeized quarks}

\author{MAXIM NEFEDOV}
%\footnote{Typeset names in
%8~pt Times Roman, uppercase. Use the footnote to indicate the
%present or permanent address of the author.}}

\address{Department of Physics, Samara National Research University, Moskovskoe Shosse, 34\\
Samara, 443086, Russia\\
nefedovma@gmail.com}
%\footnote{Typeset author e-mail
%address in single line.} }

\author{VLADIMIR SALEEV}

\address{Department of Physics, Samara National Research University, Moskovskoe Shosse, 34\\
Samara, 443086, Russia\\
saleev@samsu.ru}

\maketitle

\begin{history}
\received{Day Month Year} \revised{Day Month Year}
%\accepted{(Day Month Year)}
%\comby{(xxxxxxxxxx)}
\end{history}

\begin{abstract}
%% Text of abstract
  The technique of one-loop calculations for the processes  involving Reggeized quarks is described in the framework of gauge invariant effective field theory for the Multi-Regge limit of QCD, which has been introduced by Lipatov and Vyazovsky. The rapidity divergences, associated with the terms enhanced by $\log (s)$, appear in the loop corrections in this formalism. The covariant procedure of regularization of rapidity divergences, preserving the gauge invariance of effective action is described.  As an example application, the one-loop correction to the propagator of Reggeized quark and $\gamma Q q$-scattering vertex are computed. Obtained results are used to construct the Regge limit of one-loop $\gamma\gamma\to q\bar q$ amplitude.
  The cancellation of rapidity divergences and consistency of the EFT prediction with the full QCD result is demonstrated.
  The rapidity renormalization group within the EFT is discussed.
\end{abstract}


%% main text
\section*{Introduction.}

%% The Appendices part is started with the command \appendix;
%% appendix sections are then done as normal sections
%% \appendix

%% \section{}
%% \label{}

%% References
%%
%% Following citation commands can be used in the body text:
%% Usage of \cite is as follows:
%%   \cite{key}         ==>>  [#]
%%   \cite[chap. 2]{key} ==>> [#, chap. 2]
%%

 In the limit of (Quasi-)Multi-Regge Kinematics (MRK), partons produced in the inelastic collision of two highly-energetic strongly-interacting particles, are clustered into groups, separated by large rapidity gaps. This kinematical limit for scattering amplitudes (see~\cite{IFL, QMRKrev} for a review) is an important technical tool for the studies in different areas of Quantum Field Theory. The recent study of analytic structure of scattering amplitudes in ${\cal N}=4$ Supersymmetric Yang-Mills theory~\cite{Bartels_N4} could serve as an example of application of MRK-limit and concept of gluon Reggeization in the realm of formal theory. 
 
 On a phenomenological side, the Multi-Regge limit is relevant for the calculations of the cross-sections of production of dijets~\cite{MN_dijets} or hadron pairs~\cite{dihadr}, widely separated in rapidity. Also it serves as a theoretical basis of $k_T$-factorization~\cite{kTf} and Parton Reggeization Approach(PRA) to multiscale observables in hard processes (see e.~g.~\cite{BB_corr, diphotons, gamma-jet, dijets}).

 The Gauge-invariant effective field theory (EFT) for Multi-Regge processes in QCD has been proposed by L. N. Lipatov for the case of amplitudes with exchange of Reggeized gluons in $t$-channel~\cite{LipatovEFT}, and by L. N. Lipatov and M. I. Vyazovsky for the general case of amplitudes with Reggeized quarks and gluons in $t$-channel~\cite{LipVyaz}. This EFT is a powerful tool for studies of QCD scattering amplitudes in MRK limit.  Studies of loop corrections in this EFT has been initiated in the Refs.~\refcite{MH-ASV_NLO_qR-vert}-\refcite{MH-ASV_2loop-traj} for the case of Reggeized gluons. In the present note we perform a first exploratory study of the structure of one-loop corrections in EFT~\cite{LipVyaz} for the case of Green's functions containing the Reggeized quarks. 

 Our main concern is the problem of rapidity divergences, arising in course of calculation of loop corrections  in the formalism of Lipatov's High energy EFT~\cite{LipatovEFT, LipVyaz}.  The paper is organized as follows. In the Sec.~\ref{sec:EFT}
  the general construction of EFT for Reggeized quarks~\cite{LipVyaz} is outlined and covariant regularization scheme for rapidity
  divergences~\cite{MH-ASV_NLO_qR-vert, MH-ASV_NLO_gR-vert, MH-ASV_2loop-traj} is introduced. In the Sec.~\ref{sec:1-loop} the one-loop corrections
  to the self-energy of Reggeized quark and $\gamma Q q$-scattering vertex are computed. In the Sec.~\ref{sec:comp-QCD} the MRK-asymptotics
  for the NLO correction to the $\gamma\gamma\to q\bar{q}$-amplitude is constructed using the obtained results. After the proper
   {\it localization in rapidity} of the obtained $\gamma Q q$-vertex, the rapidity divergences in the final result cancel and the EFT result
    correctly predicts the leading-power term of the full QCD amplitude. Also a few comments concerning the Rapidity Renormalization Group are made.

\section{Effective action and rapidity divergences}\label{sec:EFT}

 In the present note we will use the following convention for the Sudakov(light-cone) decomposition for the arbitrary four-vector $k^\mu$:
 \begin{equation}
 k^\mu = \frac{1}{2} \left( n^{\mu}_+ k_- + n^{\mu}_- k_+ \right) + k_{T}^\mu,
 \end{equation}
 where the light-cone vectors $n_\pm^\mu$ are pointing along the direction of momentum of highly-energetic particles in the initial state of the collision, $n_\pm^2 = 0$, $n_+n_-=2$, $n_\pm k_T=0$. Also, in contrast to the notation of Refs.~\refcite{LipatovEFT}, \refcite{LipVyaz}, we do not distinguish the covariant and contravariant light-cone components, i.e. $k_\pm=k^\pm$ and $n^\mu_\pm=\left(n^\pm \right)^\mu$. Hence the scalar product of two four-vectors $k$ and $q$ reads:
 \[
 kq=\frac{1}{2}(k_+q_- + k_- q_+) - {\bf k}_T {\bf q}_T.
 \] 

 To set the notation, let us briefly describe the structure of gauge-invariant EFT for Multi-Regge
  processes with quark exchange in $t$-channel, introduced in the Ref.~\refcite{LipVyaz}. In this approach, the axis of rapidity
   $y=\log(q^+/q^-)/2$ is sliced into a few regions $y_i\leq y\leq y_{i+1}$, corresponding to the clusters of produced particles,
   and a separate copy of the full QCD-Lagrangian is defined in each region:
  \[
  L_{\rm QCD}(A_\mu, \psi_q)=-\frac{1}{2}{\rm tr}\left[G_{\mu\nu}G^{\mu\nu}\right]+\sum\limits_{q=1}^{n_F}\bar{\psi}_q (i\hat{D}-m_q)\psi_q,
  \]
  where $\hat{D}=\gamma_\mu D^\mu$, $D_\mu=\partial_\mu+ig_s A_\mu$ -- covariant derivative, containing the gluon field $A_\mu=A_\mu^a T^a$, where $T^a$ are (Hermitian) generators of fundamental representation of $SU(N_c)$ gauge group, $g_s$ is the Yang-Mills coupling constant and $G_{\mu\nu}=-i\left[D_\mu,D_\nu\right]/g_s$ is the non-Abelian field-strength tensor. The full effective Lagrangian for the QMRK processes with quark exchanges in $t$-channel reads:
  \begin{eqnarray}
 & L_{\rm eff}= L_{\rm kin.}(Q_+, Q_-) + \sum\limits_i \left[ L_{\rm QCD}(A_\mu^{[y_i, y_{i+1}]}, \psi_q^{[y_i, y_{i+1}]})\right. \nonumber \\
 & \left. + L_{\rm ind.}(A_\mu^{[y_i, y_{i+1}]}, \psi_q^{[y_i, y_{i+1}]}, Q_+, Q_-)  \right],   \label{eq:Leff}
  \end{eqnarray}
where the label $[y_i, y_{i+1}]$ denotes that the real part of rapidity $y$ of the momentum modes of the corresponding field is
constrained: $y_i\leq y \leq y_{i+1}$. Quarks and gluons, produced in the different intervals in rapidity, communicate via the
exchange of Reggeized quarks with the kinetic term:
\begin{equation}
L_{\rm kin.}(Q_+, Q_-)=2\left( \overline{Q}_+ i\hat{\partial} Q_- + \overline{Q}_- i\hat{\partial} Q_+ \right), \label{eq:kinQ}
\end{equation}
and the fields $Q_+$ or $Q_-$ are subject to the following kinematic constraints:
\begin{eqnarray}
  \partial_\pm Q_\mp =0,\ \partial_\pm \overline{Q}_{\mp} =0, \label{eq:kin_cons_Q1} \\
  \hat{n}_\pm Q_{\mp} =0,\ \overline{Q}_{\pm} \hat{n}_{\mp}=0. \label{eq:kin_cons_Q2}
\end{eqnarray}
  where $\partial_\pm=n_\pm^\mu \partial_\mu=2 \partial/\partial x_\mp$, $x_\pm=x^\pm=n_\pm x=x^0\pm x^3$
  in the center of mass frame of initial state.  Conditions (\ref{eq:kin_cons_Q1}, \ref{eq:kin_cons_Q2}) are equivalent
  to the requirement of QMRK for the particles in the final state. Consequently, the propagator of Reggeized quark
   $i \hat{P}_\pm \hat{q} \hat{P}_\pm/q^2$ contains the projectors $\hat{P}_\pm=\hat{n}_\mp\hat{n}_\pm /4 $ ensuring
    the constraint (\ref{eq:kin_cons_Q2}). The fields $Q_\pm$ are gauge invariant, and the corresponding {\it induced} interaction term reads:
  \begin{equation}
L_{\rm ind.}= - \overline{Q}_- i\hat{\partial} \left( W^\dagger [A_+] \psi \right) - \overline{Q}_+ i\hat{\partial} \left( W^\dagger [A_-] \psi \right) + {\rm h. c.}, \ \label{eq:Lind}
  \end{equation}
   where $W[A_\pm]=P\exp\left[\frac{-ig_s}{2} \int\limits_{-\infty}^{x_{\mp}} dx'_{\mp} A_\pm (x_{\pm},x'_{\mp},{\bf x}_T)  \right]$
   is the Wilson line stretched along the light-cone, which can be expanded perturbatively as:
  \begin{equation}
  W[A_\pm]= 1-ig_s(\partial_\pm^{-1} A_\pm)+ (-ig_s)^2 (\partial_\pm^{-1} A_\pm\partial_\pm^{-1} A_\pm)+... \
  \end{equation}
  leading to the infinite number of nonlocal interaction vertices of Reggeized quark ($Q_\pm$), Yang-mills quark ($\psi$) and Yang-Mills gluons ($A_\mu$). In particular, the Fadin-Sherman~\cite{FadinSherman, FadinBogdan} $gQq$-vertex and $ggQq$-vertex reads:
  \begin{eqnarray}
 ig_sT^a\gamma^{(\pm)}_\mu(q,k) &=& ig_s T^a \left(\gamma_\mu + \hat{q} \frac{n^\mp_\mu}{k^\mp} \right), \label{eq:gQq-vert} \\
 \gamma^{(\pm)}_{\mu\nu}(q,k_1,k_2) &=& ig_s^2 (n^\mp_{\mu_1} n^\mp_{\mu_2}) \hat{q} \left[ \frac{T^{a_1} T^{a_2}}{k_1^\mp(k_1+k_2)^\mp} + (1\leftrightarrow 2) \right], \label{eq:ggQq-vert}
  \end{eqnarray}
  where $q$ and $k_i$ are the (incoming) momenta of Reggeized quark and of gluons, respectively. The induced vertices contain nonlocal
  factors $1/k^\pm$ (eikonal propagators) which  in certain kinematics can lead to integrals, which are ill-defined in the dimensional regularization.
  To demonstrate this, let's study the scalar integral which will appear in the computation of the self-energy of the Reggeized quark:
  \begin{equation}
  B_0^{[+-]}({\bf p}_T)=\int\frac{d^D q}{i\pi^{D/2}} \frac{1}{q^2 (p-q)^2 [q^+] [q^-] },
  \end{equation}
  where, following the approach of Ref.~\refcite{MH_Pole-prescription}, we have assigned the Principal Value (PV) $i\varepsilon$-prescription to $1/q^\pm$ poles:
\begin{equation}
\frac{1}{[q^\pm]}=\frac{1}{2}\left(\frac{1}{q^\pm+i\varepsilon} + \frac{1}{q^\pm-i\varepsilon}\right),\label{eq:VP}
\end{equation}
 while the standard $i\varepsilon$-prescription for Feynman propagators is kept implicit.

 Due to constraint (\ref{eq:kin_cons_Q1}) the external momentum $p$ is purely transverse $p^+=p^-=0$.  Transferring to the integration
 over light-cone components $d^D q= dq^+ dq^- d^{D-2} {\bf q}_T/2$ and substituting the anzats $q^\pm=\xi^\pm e^{\pm y}\sqrt{|q^2+{\bf q}_T^2|}$, $\xi^+\xi^-={\rm sgn}(q^2+{\bf q}_T^2)$, one finds, that integral in rapidity is factorized:
  \begin{eqnarray}
 B_0^{[+-]} ({\bf p}_T) = \int\limits_{y_i}^{y_{i+1}} dy \int \frac{d^{d-2}{\bf q}_T}{2\pi^{D/2} i} \int\limits_{-\infty}^{+\infty} \frac{d q^2}{(q^2+{\bf q}_T^2-i\varepsilon) (q^2+i\varepsilon)}\times \nonumber \\
  \frac{1}{(q^2+2{\bf p}_T {\bf q}_T - {\bf p}_T^2 + i\varepsilon)} = \frac{(y_{i+1}-y_i)}{\pi^{(D-2)/2}} \int \frac{d^{D-2}{\bf q}_T}{{\bf q}_T^2 ({\bf p}_T-{\bf q}_T)^2}. \label{eq:B0pm_y}
  \end{eqnarray}

 The typical two-dimensional IR-divergent integral, contributing to the LO Regge trajectories of gluon~\cite{LipatovEFT} and
 quark~\cite{FadinSherman, FadinBogdan}, appears in the last expression in front of $(y_{i+1}-y_i)$-divergence.

 The regularization of EFT~\cite{LipatovEFT,LipVyaz} by explicit cutoffs in rapidity is very physically clear. The $y_i$-dependence
 should cancel between the contributions of neighboring regions in rapidity order-by-order in $\alpha_s$, building up a factor $(\log s)^n$. However, in practical applications to the multiscale Feynman integrals, the cutoff regularization is quite inconvenient. We would like to keep manifest Lorentz-covariance of the formalism and to be able to make shifts of integration momenta to follow the standard technique of tensor reduction of Feynman integrals. Also we would like to preserve the gauge invariance of EFT~\cite{LipatovEFT,LipVyaz} after regularization. Different versions of the analytic regularization for rapidity divergences can be found in the literature on Soft-Collinear Effective Theory (see e. g.~\cite{SCET_RRG}), however the proof of gauge invariance of regularized expressions is quite involved in this case. In the present study we use, {\it covariant regularization} (or regularization by {\it non light-like} or {\it tilted} Wilson lines, see e.g.~\cite{Collins}) developed in Refs.~\refcite{MH-ASV_NLO_qR-vert}-\refcite{MH-ASV_2loop-traj} for the case of Lipatov's theory~\cite{LipatovEFT}. Following this approach, we shift the $n_\pm$-vectors in the induced interactions (\ref{eq:Lind} -- \ref{eq:ggQq-vert}) slightly from the light-cone:
 \begin{eqnarray}
& n_\pm^\mu \to \tilde{n}_\pm^\mu= n_\pm^\mu + e^{-\rho} n_\mp^\mu ,\  k_\pm \to \tilde{k}_\pm= k_\pm + e^{-\rho} k_\mp ,  \label{eq:Npm} \\
& \tilde{n}^2_\pm=4 e^{-\rho},\ \tilde{n}_+\tilde{n}_-=2(1+e^{-\rho}),\label{eq:Npm2}
 \end{eqnarray}
 where $\rho\gg 1$ is the regulator variable. In the next section we will list all single-scale 1,2 and 3-point one-loop scalar integrals with one scale of virtuality, and comment on their rapidity divergences.

\section{NLO corrections to the $\gamma Q q$-vertex and Reggeized quark self-energy}\label{sec:1-loop}

  First, we list the one-loop Feynman integrals appearing in the calculation. We will work in $D=4-2\epsilon$-dimensions to regularize the
   UV and IR divergences and add the factor $r_\Gamma=\Gamma^2(1-\epsilon)\Gamma(1+\epsilon)/\Gamma(1-2\epsilon)$ to the denominators of
   all integrals to be compatible with the notation of Ref.~\refcite{K_Ellis} for ordinary scalar integrals.

  The following ``tadpole'' integral is nonzero:
  \[
  A_0^{[+]}(p)=\int \frac{[d^Dq]}{(p-q)^2[\tilde{q}^+]}=  \frac{\tilde{p}^+}{\cos (\pi\epsilon)} \frac{ e^{\rho(1-\epsilon)}}{2\epsilon(1-2\epsilon)}\left\lbrace \frac{\mu}{\tilde{p}^+} \right\rbrace^{2\epsilon}, %\label{eq:A0p}
  \]
  where $[d^Dq]= (\mu^2)^\epsilon d^Dq/(i\pi^{D/2} r_\Gamma)$ and $\left\lbrace \frac{\mu}{k} \right\rbrace^{2\epsilon}=\frac{1}{2}\left[ \left( \frac{\mu}{k-i\varepsilon} \right)^{2\epsilon} + \left( \frac{\mu}{-k-i\varepsilon} \right)^{2\epsilon} \right]$. Integral $A_0^{[+]}$ contains the {\it power divergence} $\sim e^\rho$. Here and in the following, the integral $I^{[-]}$ can be obtained by the $(+\leftrightarrow -)$ substitution in the $I^{[+]}$ integral.

  The asymptotic expansion for $\rho\gg 1$ of the following ``bubble'' integral was computed using Mellin-Barnes representation~\cite{Smirnov}:
  \[
  B_0^{[+]}(p)=\int \frac{[d^Dq]}{q^2 (p-q)^2[\tilde{q}^+]} =  \frac{1}{\tilde{p}^+} \frac{e^{-\epsilon\rho}}{ \cos (\pi \epsilon)} \frac{1}{2\epsilon^2} \left\lbrace \frac{\mu}{\tilde{p}^+} \right\rbrace^{2\epsilon},
  \]
  and all terms $\sim e^{-n\rho/2}$ for $n\geq 1$ are neglected. In course of calculation, the leading $e^{\rho/2}$ power divergence,
  which can be found in the Ref.~\refcite{MH-ASV_NLO_gR-vert}, has canceled due to the PV-prescription (\ref{eq:VP}). This integral contains the factor $e^{-\epsilon\rho}\to 0$ for $\epsilon>0$ and $\rho\to \infty$. For the case $p^+=p^-=0$
  integral $B_0^{[+]}({\bf p}_T)=0$ because Mellin-Barnes representation starts with terms $\sim e^{-\rho/2}$.

  In course of tensor reduction, one has to do the shifts of loop momenta of the type: $(q-p)^2(q-k)^2\to q^2 (k+p-q)^2$, then the following integral appears:
  \[
  \int\frac{[d^Dq] }{q^2(p+k-q)^2 [\tilde{q}^+-\tilde{p}^+]}=B_0^{[+]}(p+k)-\frac{\tilde{p}^+}{\tilde{k}^+} B_0^{[+]}(p),
  \]
  where $\tilde{p}^+\sim e^{-\rho}$ in our calculation.

  The integral with two eikonal propagators:
  \begin{equation}
  B_0^{[+-]}({\bf p}_T)=\frac{1}{{\bf p}_T^2}\left(\frac{\mu^2}{{\bf p}_T^2}\right)^\epsilon \frac{i\pi - 2\rho}{\epsilon},
  \end{equation}
  contains the {\it logarithmic} rapidity divergence $\sim \rho$ and can be extracted from the results of Ref.~\refcite{MH-ASV_NLO_qR-vert}.
  It is easy to see, that logarithmic divergence $\sim\rho$ is equivalent to $(y_{i+1}-y_i)$ in Eq. (\ref{eq:B0pm_y}), and also the imaginary
   part appears, due to PV prescription (\ref{eq:VP}).

  The ``triangle'' integral with one eikonal propagator:
  \[
  C_0^{[+]}(k^+,{\bf p}_T^2)=\int \frac{[d^Dq]}{q^2 (p-q)^2 (p+k-q)^2 [\tilde{q}^+]},
  \]
  where $k^2=0$, $k^-=0$, $p^2=-{\bf p}_T^2$, $(p+k)^2=0$ is obtained using Mellin-Barnes representation. Our result coincides
   with the result of Ref.~\refcite{MH-ASV_NLO_gR-vert}:
   \begin{eqnarray*}
  C_0^{[+]}(k^+, {\bf p}_T^2) = \frac{1}{k^+ {\bf p}_T^2}\left( \frac{\mu^2}{{\bf p}_T^2} \right)^\epsilon \frac{1}{\epsilon} \left[  \rho -i\pi + \log \frac{(k^+)^2}{{\bf p}_T^2} +  \psi(1+\epsilon) + \psi(1) - 2\psi(-\epsilon) \right],
   \end{eqnarray*}
   and contains the logarithmic rapidity divergence and imaginary part, similarly to the integral $B_0^{[+-]}$.  These are all one-loop rapidity-divergent
    integrals with one virtuality scale, which appear in EFT~\cite{LipatovEFT,LipVyaz}.

   The tensor reduction of one-loop integrals follows the usual Passarino-Veltman procedure, except that vectors
    $\tilde{n}_\pm$ should be added to the anzats for the tensor structure. For example, solution for the integral $B^{[+]}(p)$
    with one factor of $q_\mu$ in the numerator reads:
\begin{eqnarray*}
  \int \frac{\left[ d^Dq \right] q^\mu}{q^2 (p-q)^2 [q^+]} &=& \frac{1}{\Delta} \left\{ \left[ \tilde{p}^+ B_0(p^2) - 2e^{-\rho} \left(A_0^{[+]}(p) + p^2 B_0^{[+]}(p)\right) \right] p^\mu  \right. \\
 && \left. + \left[ \tilde{p}^+ \left( A_0^{[+]}(p) + p^2 B_0^{[+]}(p) \right) - 2 p^2 B_0(p^2) \right] \tilde{n}_+^\mu \right\},
\end{eqnarray*}    
   where $\Delta=(\tilde{p}^+)^2-4p^2 e^{-\rho} $ is the Gram determinant and $B_0(p^2)$ is the usual one-loop massless bubble integral~\cite{K_Ellis}. In the case when  $\tilde{p}^+\sim e^{-\rho}$,
     the Gram determinant will be proportional to $e^{-\rho}$ and the overall result will contain power divergence $\sim e^\rho$. Because of this power divergences one has to keep
     the exact $\rho$-dependence of coefficients in front of scalar integrals, arising from Eq.~(\ref{eq:Npm2}), up to the moment when
     all power divergences cancel out. Then, the terms suppressed by powers of $e^{-\rho}$ can be dropped.

   \begin{figure}[ht]
   \begin{center}
   \begin{tabular}{ccc}
   \includegraphics[width=0.70\textwidth]{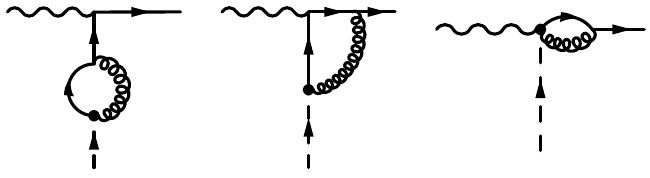} &
   \includegraphics[width=0.14\textwidth]{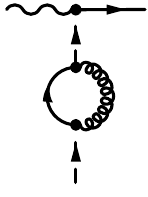} \\
   (a) & (b)
   \end{tabular}
   \caption{Diagrams, contributing to $\gamma Q q$-vertex at one-loop -- (a) and subtraction term for the one-loop $\gamma Qq$-vertex -- (b).
   Reggeized quark is denoted by dashed line.   \label{fig1}}
   \end{center}
   \end{figure}

   Now we are in a position to present the results of computation of self-energy of the Reggeized quark and the $\gamma Qq$-vertex at one-loop.
   The one-loop Reggeized quark self-energy reads:
   \begin{eqnarray*}
  \hat{\Sigma}_1({\bf p}_T) &=& \frac{C_F\bar{\alpha}_s}{4\pi} (-i)\int [d^Dq] \frac{\gamma_\mu^{(-)}(-p,q) (\hat{p}-\hat{q}) \gamma_\mu^{(+)}(p,-q)}{q^2(p-q)^2},
   \end{eqnarray*}
 where $\bar{\alpha}_s=\alpha_s (4\pi/\mu^2)^\epsilon r_\Gamma$ and $C_F=(N_c^2-1)/(2N_c)$. And the result is:
   \begin{eqnarray}
  \hat{\Sigma}_1({\bf p}_T)=(-i\hat{p}) \frac{C_F \bar{\alpha}_s}{4\pi} \left(\frac{\mu^2}{{\bf p}_T^2}\right)^\epsilon \left[ \frac{{2\rho}-i\pi}{\epsilon} + \left(\frac{1+\epsilon}{1-2\epsilon}\right) \frac{1}{\epsilon} \right].
   \end{eqnarray}
   Apart from the $\rho$-divergence, leading to the quark Reggeization, the self-energy graph also contains UV-pole.
   The integrals containing power rapidity divergences do not appear in this calculation.

   Feynman diagrams contributing to the one-loop correction to $\gamma Qq$-vertex are shown in the Fig.~\ref{fig1}(a).
   We will consider the kinematics, when high-energy photon with large $k^+$ momentum component scatters on the Reggeized
   quark with virtuality $p^2=-t_1$, $t_1>0$ and produces massless quark. All scalar integrals, described above, appear in the result of tensor reduction,
   but the coefficients in front of integrals $A_0^{[+]}$ and $B_0^{[+]}$ are of order $O(e^{-2\rho})$ or $O(e^{-\rho})$, and,
   therefore, the result is free from power rapidity divergences. To obtain the result free from the terms $\sim e^{\pm \epsilon \rho}$ one have to take into account the corresponding high-energy projector $\hat{P}_\pm$ in the propagator of Reggeized quark. Thus, only logarithmic rapidity divergence, originating from integral $C_0^{[+]}$ is left. After neglecting all terms suppressed by powers of $e^{-\rho}$ one can expand the result
   in $\epsilon$. The expanded result reads:
   \begin{eqnarray}
  \hat{\Gamma}_1^\mu=\frac{2}{t_1}\hat{\Delta}_0^\mu +\hat{\Gamma}_0^\mu\left[-\frac{1}{\epsilon^2}-\frac{L_1}{\epsilon} +(\rho-i\pi) \left(\frac{1}{\epsilon}+L_1\right) +\right. \nonumber \\
 \left. \frac{2L_2}{\epsilon} - \left(\frac{1}{\epsilon}+L_1+3\right) +2L_1L_2 - \frac{L_1^2}{2} + \frac{\pi^2}{2}  \right]  + O(\epsilon),\label{eq:unsVert}
   \end{eqnarray}
   where the overall factor $\bar{\alpha}_sC_F/(4\pi)$ is omitted, $L_1=\log\left({\mu^2}/{t_1}\right)$, $L_2=\log\left({k^+}/{\sqrt{t_1}}\right)$ and the Dirac structures $\hat{\Gamma}_0$ and $\hat{\Delta}_0$ have the form:
   \begin{eqnarray*}
   \left(\hat{\Gamma}_0 \right)_\mu &=& i ee_q \bar{u}(p+k)\gamma^{(-)}_\mu(p,k)\hat{P}_-,\\
   \hat{\Delta}_0^\mu &=& i ee_q \left(p^\mu - \frac{t_1}{2k^+} n_+^\mu\right) \left( \bar{u}(p+k) \hat{k} \hat{P}_- \right),
   \end{eqnarray*}
   where $e$ is the QED coupling constant and $e_q$ is the quark electric charge. It is important to notice, that the obtained result
   satisfies the QED Ward identity $k_\mu \hat{\Gamma}_1^\mu=0$ independently on $t_1$. In fact, the full $\rho$-dependent result, before
   omitting the power-suppressed terms, also obeys Ward identity, and the third diagram in the Fig.~\ref{fig1}(a) is particularly
   important for this. In Ref.~\refcite{MH-ASV_NLO_gR-vert} contributions of diagrams of this topology where nullified by the gauge-choice
   for external gluons.

  Covariant rapidity regulator breaks-down explicit separation of the contributions of different regions in rapidity.
  Regularization of the $1/{q^+}$-pole in the vertex correction corresponds to the cutoff of the loop momentum at some
  large negative rapidity $y_1(\rho)<0$, while regularization of $1/q^+$ and $1/q^-$-poles in the self-energy correction
  constrains the rapidity of gluon in the loop from both sides: $y_1(\rho)<y<y_2(\rho)$. Clearly, if we just add the Reggeon
  self-energy graph to the vertex correction, the contribution of region $y_1(\rho)<y<y_2(\rho)$ will be double-counted and there
   is no reason to expect the cancellation of $\sim\rho$ divergences. Before adding all NLO corrections together, one have to ``localize''
   the vertex correction in rapidity by subtraction of the corresponding central-rapidity
   contribution~\cite{MH-ASV_NLO_qR-vert, MH-ASV_NLO_gR-vert, MH-ASV_2loop-traj}. The diagrammatic representation for the corresponding
    subtraction term is shown in the Fig.~\ref{fig1}(b), and after $\epsilon$-expansion it has the form:
  \begin{equation}
  \delta\hat{\Gamma}_1^\mu=\hat{\Gamma}_0^\mu\left[ (2\rho-i\pi) \left(\frac{1}{\epsilon}+L_1 \right) + \left( \frac{1}{\epsilon}+L_1+3 \right)\right] + O(\epsilon). \label{eq:sub_term}
  \end{equation}
  Apart from the rapidity divergence, the subtraction term contains also $1/\epsilon$ UV-pole, associated with the UV-divergence
   of the ordinary quark propagator.

\section{Comparison with QCD. Rapidity Renormalization Group.}\label{sec:comp-QCD}
  The results obtained above can be used to construct the Regge asymptotics of $\gamma\gamma\to q\bar{q}$ scattering amplitude at one loop.
  To facilitate the comparison with the full QCD result we will consider the scalar quantities -- the real and imaginary parts of the interference
  term between LO tree-level and one-loop amplitude. In the Regge limit, the parameter $\tau=-t/s$ is small, and the interference term can be
  expanded as follows:
  \begin{equation}
  \sum\limits_{\rm col.,\ spin} M_{\rm 1-loop} M^*_{\rm LO}  = \frac{1}{\tau}C_{HE}^{(-1)} + C_{HE}^{(0)} + O(\tau),
  \end{equation}
  where the overall factor $(8\cdot (e e_q)^4 N_c)\frac{C_F \bar{\alpha}_s}{4\pi}$ is omitted and the EFT~\cite{LipVyaz} predicts the coefficient
   in front of the leading power -- $C_{HE}^{(-1)}$.

   After the subtraction of the contribution (\ref{eq:sub_term}) from the corrections to the $\gamma Q_+ q$ and $\gamma Q_- q$ scattering vertices,
   all rapidity divergences cancel at the order $\alpha_s$, and the results for the real and imaginary parts of the coefficient $C_{HE}^{(-1)}$
   looks as follows:
   \begin{eqnarray}
  {\rm Re}\ C_{HE}^{(-1)} &=&-\frac{2}{\epsilon^2} - \frac{2}{\epsilon} \log \frac{\mu^2}{(-t)}+ \left( 1+2\log\frac{1}{\tau} \right)\left(\frac{1}{\epsilon}+\log \frac{\mu^2}{(-t)}\right)\nonumber\\
&&- \left[ 3-\pi^2 +\log ^2 \frac{\mu^2}{(-t)}+4\log\frac{1}{\tau}\right] , \label{eq:CHE_re}\\
  {\rm Im}\ C_{HE}^{(-1)}&=&-\pi \left( \frac{1}{\epsilon} + \log \frac{\mu^2}{(-t)}  -2  \right) . \label{eq:CHE_im}
   \end{eqnarray}

\begin{figure}
\begin{center}
\begin{tabular}{cc}
\includegraphics[width=0.34\textwidth]{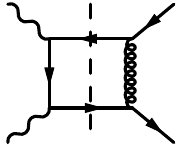} & \includegraphics[width=0.34\textwidth]{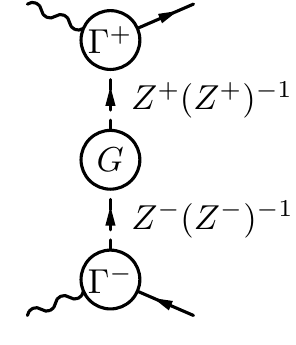} \\
(a) & (b)
\end{tabular}
\end{center}
\caption{The cut contributing to the imaginary part of $\gamma\gamma\to q{\bar q}$ amplitude -- (a).
Contribution with the exchange of one Reggeon in $t$-cahnnel -- (b).\label{fig2}}
\end{figure}

   We have performed the computation of real and imaginary parts of $C_{HE}^{(-1)}$ in full QCD in dimensional regularization,
   using the \texttt{FeynArts}~\cite{FeynArts} and \texttt{FeynCalc}~\cite{FeynCalc} packages. The obtained results agree with
   the expressions (\ref{eq:CHE_re}), (\ref{eq:CHE_im}), derived within the EFT~\cite{LipVyaz}, providing the nontrivial test
   of the self-consistency of the formalism. In particular, the imaginary part (\ref{eq:CHE_im}) originates from the box diagram
   with quark and gluon exchange in the $t$-channel (Fig.~\ref{fig2}(a)).

   The EFT~\cite{LipatovEFT,LipVyaz} provides an efficient tool for resummation of high-energy logarithms $\sim\log 1/\tau$. Assuming,
   that after appropriate localization of all loop corrections in rapidity, logarithmic rapidity divergences cancel order by order in $\alpha_s$
   within the subset of diagrams with exchange of one Reggeon in $t$-channel (Fig.~\ref{fig2}(b)), one can introduce the multiplicative
   renormalization to remove them from corrections to the $\gamma Q_\pm q$-vertices and propagator of the Reggeized quark~\cite{MH-ASV_2loop-traj}:
  \begin{eqnarray*}
& \Gamma^\pm_R\left(\frac{k^\pm}{M^\pm}, \frac{t}{\mu^2} \right)= Z^\pm\left(\rho, \frac{t}{\mu^2} ,\frac{M^\pm}{\sqrt{\mu^2}}\right) \Gamma^\pm\left( \rho, \frac{k^\pm}{\sqrt{\mu^2}}, \frac{t}{\mu^2} \right),\\
& G_R\left(\frac{M^+}{\sqrt{\mu^2}}, \frac{M^-}{\sqrt{\mu^2}}, \frac{t}{\mu^2} \right) = (Z^+Z^-)^{-1} G\left(\rho, \frac{t}{\mu^2} \right),
\end{eqnarray*}
  where the scales $M^\pm$ are introduced to parametrize the finite ambiguity in the definition of subtraction scheme for $\rho$-divergences,
  related with the possible redefinition of the regulator variable, which doesn't spoil the $t$-channel factorization of the amplitude: $\rho \to \rho - 2\log M^\pm/\sqrt{\mu^2}$. The bare quantities do not depend on $M^\pm$, and therefore the renormalized quantities obey the Rapidity Renormalization Group (RRG) equations:
    \[
  \frac{\partial\log G_R}{\partial\log M^\pm}=\omega(t),\  \frac{\partial\log \Gamma^\pm_R}{\partial\log M^\pm}=-\omega(t),
  \]
  where the corresponding anomalous dimension  $\omega(t)=\lim\limits_{\rho\to\infty} \frac{(-1)}{Z^\pm}\frac{\partial Z^\pm}{\partial \log M^\pm}$
  is nothing but the quark Regge trajectory. At one loop we have:
  \begin{eqnarray*}
 Z^\pm &=& 1+ \frac{\bar{\alpha}_s}{4\pi}C_F \left(\rho - 2\log \frac{M^\pm}{\mu}\right)\left(\frac{1}{\epsilon}+\log\frac{\mu^2}{(-t)}\right)+O(\bar{\alpha}_s^2), \\
\omega(t) &=&\frac{\bar{\alpha}_s}{2\pi} C_F \left( \frac{1}{\epsilon}+ \log\frac{\mu^2}{(-t)} \right) + O(\bar{\alpha}_s^2,\epsilon),
  \end{eqnarray*}
  in full consistency with the results present in the literature~\cite{FadinSherman, FadinBogdan}.

  To remove the large logarithms $\log k^\pm/M^\pm$ from vertex corrections, one can set $M^+=k_1^+$ and $M^-=k_2^-$,
  then it is possible to resum the corrections enhanced by $\log (M^+M^-)= \log s$ by solving the RRG equation for the Reggeon propagator:
  \[
  G_R(t,s)=\left(\frac{s}{M_0^+M_0^-}\right)^{\omega(t)} G_R(t,M_0^+M_0^-),
  \]
  where one can take the Born propagator of the Reggeized quark as the initial condition for the evolution $G_R(t,M_0^+M_0^-)$
  at the starting scale $M_0^+M_0^-\sim -t$. In this way, the Reggeization of quark, as well as of a gluon, can be understood within
  the context of EFT~\cite{LipatovEFT,LipVyaz} with covariant rapidity regulator.

\section*{Conclusions}

  The results obtained above show, that the gauge invariant effective action for high energy processes in QCD~\cite{LipatovEFT, LipVyaz}
  together with the covariant rapidity regulator, proposed in
  the Refs.~\refcite{MH-ASV_NLO_qR-vert}-\refcite{MH-ASV_2loop-traj} and \refcite{MH_Pole-prescription} is a powerful tool for the studies of MRK asymptotics of QCD amplitudes and resummation of large logarithmic corrections enhanced by the high energy logarithms $\log s$ or $\log 1/x$ in Leading Logarithmic Approximation and beyond.

\section*{Acknowledgments}

  Authors are grateful to B. A. Kniehl, G.~Chachamis and A. Sabio Vera for clarifying communications concerning
  the computation of scalar integrals. The work was funded by the Ministry of Education and Science of
Russia under Competitiveness Enhancement Program of Samara
University for 2013-2020, project 3.5093.2017/8.9.

\end{document}